# Design of Novel 3T Ternary DRAM with Single Word-Line using CNTFET

Zarin Tasnim Sandhie, *Student Member, IEEE*, Farid Uddin Ahmed, *Student Member, IEEE*, and Masud H. Chowdhury, *Senior Member, IEEE*

*Abstract*— Ternary logic system is the most promising and pursued alternate to the prevailing binary logic systems due to the energy efficiency of circuits following reduced circuit complexity and chip area. In this paper, we have proposed a ternary 3-Transistor Dynamic Random-Access Memory (3T-DRAM) cell using a single word-line for both read and write operation. For simulation of the circuit, we have used Carbon-Nano-Tube Field Effect Transistor (CNTFET). Here, we have analyzed the operation of the circuit considering different process variations and showed the results for write delay, read sensing time, and consumed current. Along with the basic DRAM design, we have proposed a ternary sense circuitry for the proper read operation of the proposed DRAM. The simulation and analysis are executed using the H-SPICE tool with Stanford University CNTFET model.

## I. INTRODUCTION

The binary or Boolean logic system is the most prevailing logic system which is used extensively in the most computing system. The binary logic system is the base 2 logic system where every logic bit can have two possible values (0 or 1, low or high). The critical issue associated with the binary logic-based computing system is the interconnect limitation in the nanoscale range. The interconnects in integrated circuit design introduce noise, delay, and an increase in power consumption [1]. As the continuous scaling of the semiconductor industry is going on, the interconnect has become one of the key factors to define the performance of integrated circuits [2]. The use of Multiple-Valued Logic or MVL holds a great promise to address this critical issue. MVL (base 3 or more) is the logic system where each digit can hold three or more data/ information. Over the last few decades, MVL and its' applications have been investigated greatly because of the capability of the MVL logic devices to deliver a considerably higher information density along with smaller logic gates and diminished circuit complexity compared to the binary logic system. Therefore, the energy consumption, area, and circuit overheads, and other expenses for every bit of information decrease in the MVL system [3].

Multi-Valued Logic or MVL can be labeled as ternary (base 3), quaternary (base 4), quinary (base 5), and so on. Among them, ternary logic is the most sought one. Ternary logic is the most feasible logic system after binary because of the simplicity and similarity to binary logic. In real-life applications, decimal (base-10) and binary number systems are mostly used by humans and machines. But theoretically, it can be proved that the base 3 number system is the most efficient number system [4], [25].

There are different scopes available where MVL has been tried, such as arithmetic circuits, memory circuits, quantum computing, high-speed signaling etc. Till today, the direct or indirect approach of MVL can be found in different sector such as network routing [5] for IPv4 and IPv6, Ternary Content-Addressable Memories (TCAMs) [6] which are generally designed with conventional SRAM or DRAM, large flash memories having multiple bits per cell [7], wireless and wireline signal carrying more than one bit of information using QAM (Quadrature Amplitude Modulation) and QPSK (Quadrature Phase Shift Keying) etc.

Since 1974, prevailing MOSFET technologies have been used to implement ternary logic. Due to short channel effects such as scaling limits, DIBL (Drain Induced Barrier Lowering), energy consumption, and other signal integrity issues, the conventional technologies are becoming less appealing to the MVL system. Newer device technologies like Carbon Nano Tube Field Effect Transistor (CNTFET), Graphene-Nano-Ribbon Field Effect Transistor, Memristor etc., are being investigated to lessen the effect of short channel effects and design effective MVL circuits [8]-[12].

CNTFET possesses many potential benefits, for example, ballistic transport of charge carriers, low off current, controllable threshold voltage, etc. [1]. Many recent works of literature are available which work on the application of CNTFET in designing different arithmetic and memory circuits. CNTFET has gained remarkable attention in the field of MVL due to the ability of controllable threshold voltage. However, CNTFET comes with various concerns that need to be addressed before the benefits can be completely utilized in fabricated devices. A few of the important issues with CNTFET are the difficulty of mass production, avalanche breakdown at high temperature, higher production cost, etc. [13], [14]. Different successful attempts have been taken for the fabrication of CNTFET [15]-[17].

In this paper, we have proposed a ternary 3T Dynamic Random-Access Memory or DRAM cell using a 16nm

Stanford University model of CNTFET [18]. In the proposed design, we have used three CNTFETs along with a single word line for both read and write operation. Also, we have examined the delay and power consumption in the presence of various process variations using Monte Carlo simulations. Along with these, we have proposed a sense circuitry containing an enable signal which is capable of disconnecting the whole sense circuit from the main circuit upon necessity and save valuable power.

The paper organizes as follows. Section II briefly illustrates the working principle of a CNTFET. Section III presents the proposed DRAM design. Section IV represents the simulation results. Section V explains the proposed sense circuit and the simulation results associated with it. And Section V concludes the paper.

## II. CARBON NANO TUBE FIELD EFFECT TRANSISTOR

Carbon Nano Tube (CNT) is a tube made of graphene. Graphene is a two-dimensional sheet of carbon atoms closely packed into a honeycomb lattice. The electrical properties of CNT is controlled by the rolling angle of the graphene sheet, known as chirality (Fig. 1).

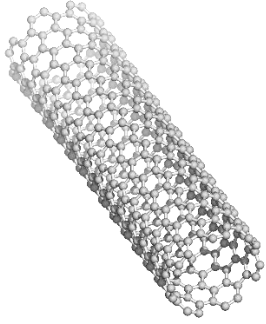

Fig. 1: Structure of a chiral CNT [20]

The chirality vector is defined by two variables: *n*, and *m*. Based on the value of the chirality vector, a CNT can act as a metal or semiconductor. For the condition where, n = m or n-m = 3i (i being an integer), the CNT exhibits metallic behavior and for n-m > 3i the CNT exhibits semiconducting behavior. For a single-walled CNT (SWCNT), the equation between the chirality vector (m, n) and tube diameter ($D_{cnt}$) can be denoted by (1).

$$D_{cnt} = \frac{a\sqrt{n^2 + m^2 + nm}}{\pi} \quad (1)$$

Here, a = 2.49 Å is the interatomic distance between two adjoining atoms. The chiral angle can be expressed by (2) [19].

$$\theta = \tan^{-1}\frac{\sqrt{3}n}{2m + n} \quad (2)$$

The bandgap of a CNT is directly dependent on the chiral angle and diameter. A CNT is free of any boundary scattering due to the absence of boundaries in a perfect and hollow cylindrical structure. Also, they are quasi-1D materials which allows only forward and backscattering. In a Carbon Nano Tube Field-Effect Transistor (CNTFET) the channel is made up of a single CNT or an array of CNTs instead of bulk silicon in the conventional MOSFET. Fig. 2 shows the structural diagram of a CNTFET. The threshold voltage ($V_{th}$) of an intrinsic CNT channel can be estimated to the first order of the half bandgap ($E_g$), which varies inversely with the tube diameter ($D_{cnt}$). The threshold voltage is modified by adjusting the diameter and is expressed by (3) [21].

$$V_{th} \approx \frac{Eg}{2e} = \frac{aV\pi}{\sqrt{3} \times eD_{cnt}} \quad (3)$$

The parameter Vπ (≈3.033 eV) is the energy of the carbon π-π bond in the tight-bonding model, *e* is the charge of a single electron. If the value of *m* in the chirality vector is considered to be 0, it can be obtained from (1) and (3) that the threshold voltage of the device is inversely proportional to the chirality vector *n*.

$$\frac{V_{th1}}{V_{th2}} = \frac{D_{cnt2}}{D_{cnt1}} = \frac{n_2}{n_1} \quad (4)$$

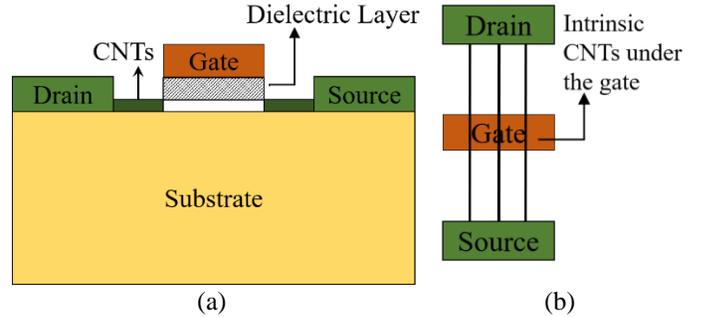

Fig. 2: Cross section view (b) Top view of a CNTFET structure

## III. PROPOSED DESIGN OF TERNARY DRAM

The proposed diagram of ternary 3T-DRAM is given in Fig. 3. Here, one word-line (WL) and two bit-lines (BL) has been used for the proper operation of the DRAM. The control signal is given by WL. Depending on the value of WL, the DRAM cell operates in a write or read mood. BLs are used for the reading or writing operation of data.

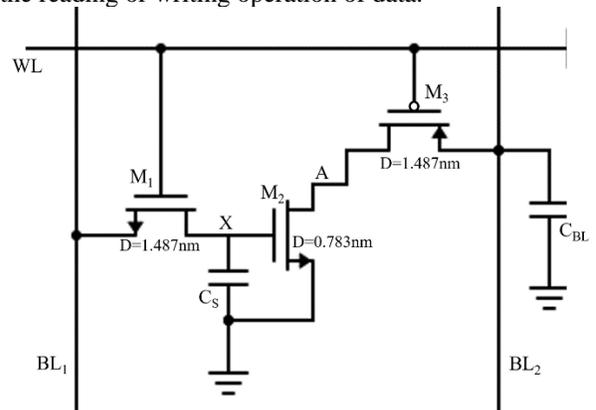

Fig. 3: Schematic diagram of the proposed 3T DRAM

BL1 takes part in the write operation and BL$_2$ takes part in the read operation of the memory cell directly. The threshold voltage of the three transistors M$_1$, M$_2$, and M$_3$ are 0.24V, 0.6V, and -0.24V, respectively. Here, D denotes the CNT diameter used in the transistors. Two capacitors C$_S$ and C$_{BL}$ are used for the data storage and discharge operation. The values of the capacitors are 0.1fF and 0.7fF, respectively.

*A. Write Operation*

The data that needs to be written in the memory cell is placed in BL$_1$. Upon the arrival of the positive edge of the word-line WL, the transistor M$_1$ starts conducting. As a result, the data starts to store in the capacitor C$_S$. For storing the data "1" or 0.5V$_{DD}$, the exact 0.5V$_{DD}$ gets stored in the capacitor. For the storage of the data "2" or V$_{DD}$, a threshold voltage drop happens like the binary DRAM. Once the data is written in the node "X", it stays there until the next positive edge of the word-line.

*B. Read Operation*

When the value on the word-line becomes zero, the read operation begins. At that time, the transistor M$_3$ turns on. The bit-line BL$_2$ is pre-charged to the voltage V$_{DD}$. If the data stored in X is zero, then transistor M$_2$ remains off and the voltage at node A remains high. If the data stored in the cell is "1" or "2", then M$_2$ turns on, and BL$_2$ discharges through the following path. The rate of discharge is different for the read operation of "1" and "2", which is shown in Fig. 6. This is an inverting memory cell for which the inverted value of the data is obtained from the BL$_2$.

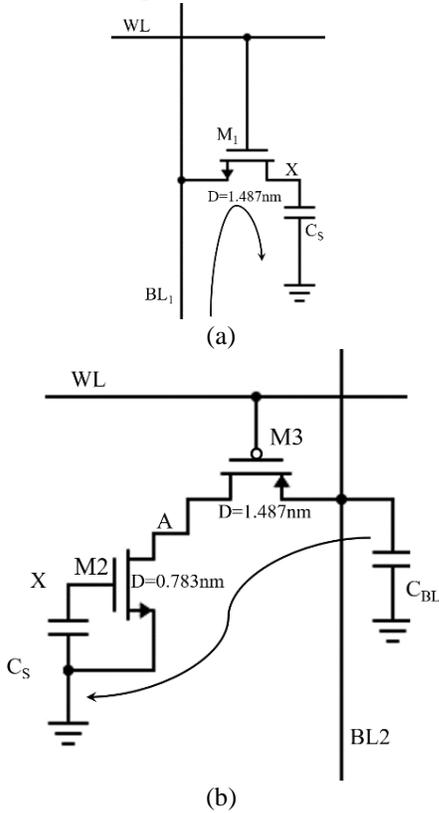

Fig. 4: (a) Write (b) Read operation of 3T DRAM

## IV. SIMULATION RESULTS AND ANALYSIS

Different parameter values that have been used during the simulation are given in Table 1. Fig. 5 shows the transient response of the proposed DRAM cell for a single cycle. For three consecutive cycle of WL, three different data ("0", "1" and "2") have been written and read. For the first cycle, "0" is stored in X and then in the falling edge of the WL, the bit-line holds its' pre-charged value. At the second rising edge of the word-line, write bit-line BL$_1$ is holding the value "1" which is read in the next falling edge of the WL by the BL$_2$. Similarly, at the third rising edge of the word-line, write bit-line BL$_1$ is holding the value "2" which is read in the next falling edge of the WL by the BL$_2$. The difference between reading "1" and "2" can clearly be differentiating from the slope of the discharging curve.

Table I: Different process and device parameter values

| Temperature | 25ºC |
|---|---|
| Supply voltage | 1.2V |
| Channel length | 16nm |
| Oxide Thickness | 4nm |
| Propagation delay | 50ps |

Table II: Simulation results for the ternary 3T DRAM for nominal conditions

| Write time | 31.776ps |
|---|---|
| Read sensing time | 0.55675ns |
| Total Current | 28.319nA |
| Total Power | 32.2674nW |

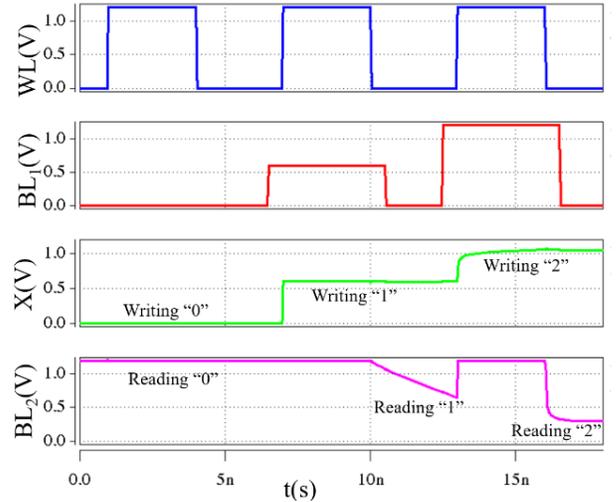

Fig. 5: Transient response of proposed DRAM showing the write and read operation of "0", "1" and "2"

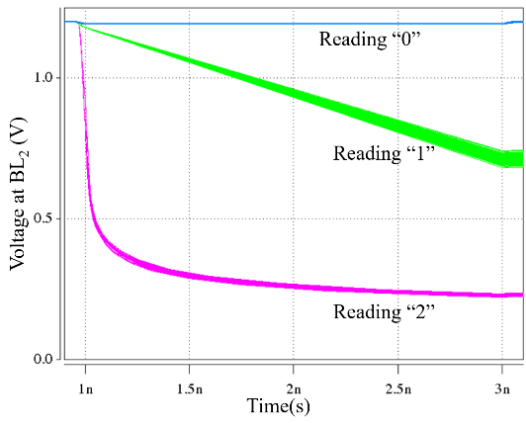

Fig. 6: Read operation of the ternary DRAM for data "0", "1" and "2" following 100MC operation with different process variation

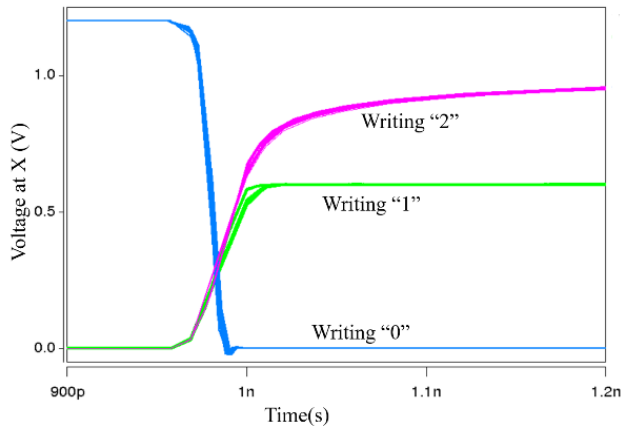

Fig. 7: Write operation of the ternary DRAM for data "0", "1" and "2" following 100MC operation with different process variation

Fig. 6 and Fig. 7 show the graphs of the reading and writing operation using 100 Monte Carlo simulations considering temperature, supply voltage, channel length, and oxide thickness variation. For the write operation, we have considered 2→0, 0→1, and 0→2 transitions. We can see from the graph that the transition between different states is most vulnerable in the state "1".

The simulated result such as write time, data sense time during read operation, and the total current are given in Table 2. These data are extracted considering the mentioned parameter values in Table 1. The write time or write delay is measured as the time required for the X node to reach to 50% value after the signal WL reaches 50% of its' value. And the read sensing time is dependent on the capacitor to be discharged to a certain value so that a sense circuit can sense the transition and decide. For that, we have calculated the read sensing time or read time as the time required by $BL_2$ to reach 20% of the data to be read. The total current is calculated as the current flowing from $BL_1 \to X$ and $BL_2 \to X$ for a total read/write cycle as shown in Fig. 5.

In Fig. 8, the effect of different parameter variations on the writing time, read sensing time, and consumed current using Monte Carlo simulations are given. Fig. 8(a)-(c) shows the effect of temperature on the write time, read time, and total current. It can be observed from the graphs that, the delay of the circuit with increasing temperature and consumed current increases with increasing temperature. Fig. 8(d)-(f) shows the effect of channel length, oxide thickness, and supply voltage variation on the calculated parameter. For the simulation, we have varied the parameters using a Gaussian function with an absolute variation for -3σ to +3σ. It can be concluded from the graphs that the effect of supply voltage variation is most dominant on the circuit outputs. For our simulation, the write time, read time and total current deviates from original value to a maximum of 5.4ps, 0.25ns, and 3.3nA for the worst case.

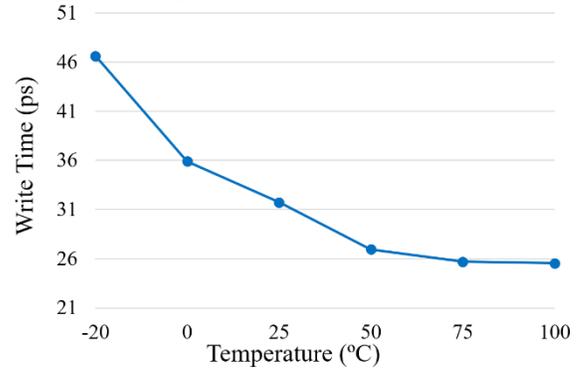

(a)

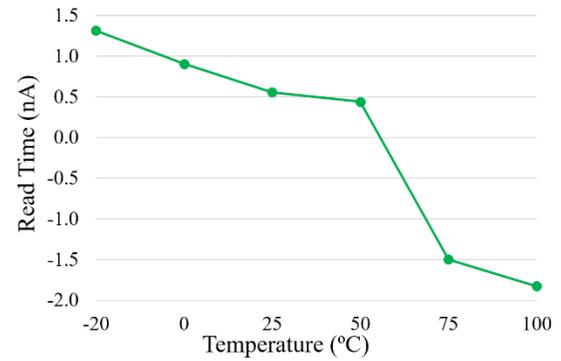

(b)

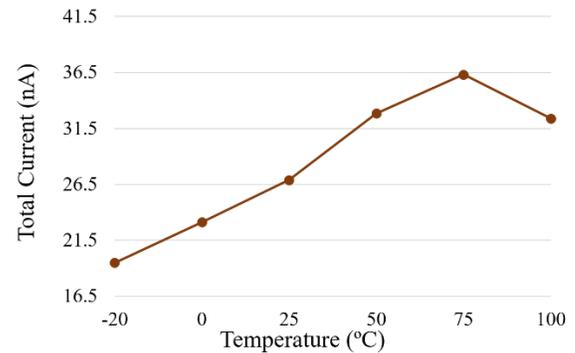

(c)

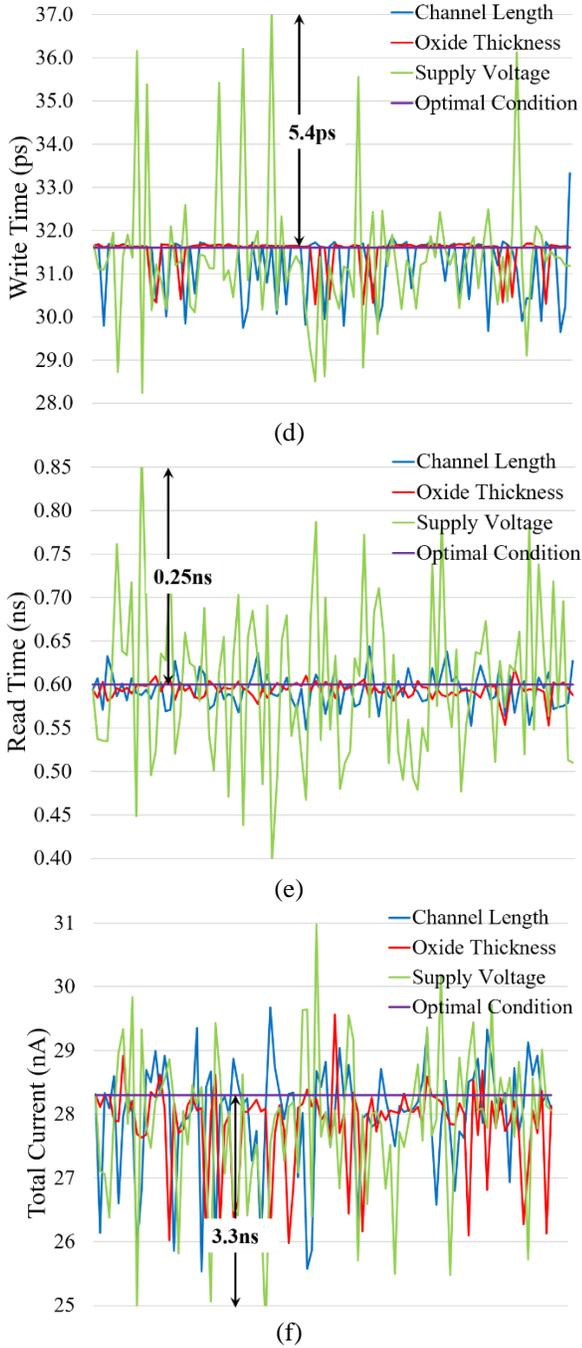

Fig. 8: Effect of temperature, supply voltage, oxide thickness, and channel length variation on write time, read sensing time, and total consumed current for the proposed design

## V. SENSE CIRCUITRY

One crucial and vital factor of a functional DRAM read circuitry is the sense amplifier. Its function is to sense the small voltage swing during reading of different logic levels and amplify the small voltage swing to a noticeable amount of voltage level. From Fig. 5, we can see that during the read operation, reading of data "0" is determined if the voltage level at the bitline BL2 remains at a higher value ($V_{DD}$). And the reading of data "1" and "2" can be distinguished from the transitional slope of the voltage level at BL$_2$. To detect these different logic levels, we have established a simple circuitry which can detect these operations successfully and generate appropriate voltage levels at the output.

Fig. 9 demonstrates the proposed sense circuitry for sensing the read signal received from BL$_2$. For the sensing purpose, we have used a ternary inverter which converts the input signal according to Table 3. A detail description of the working principle of a ternary inverter is available in [23], [24].

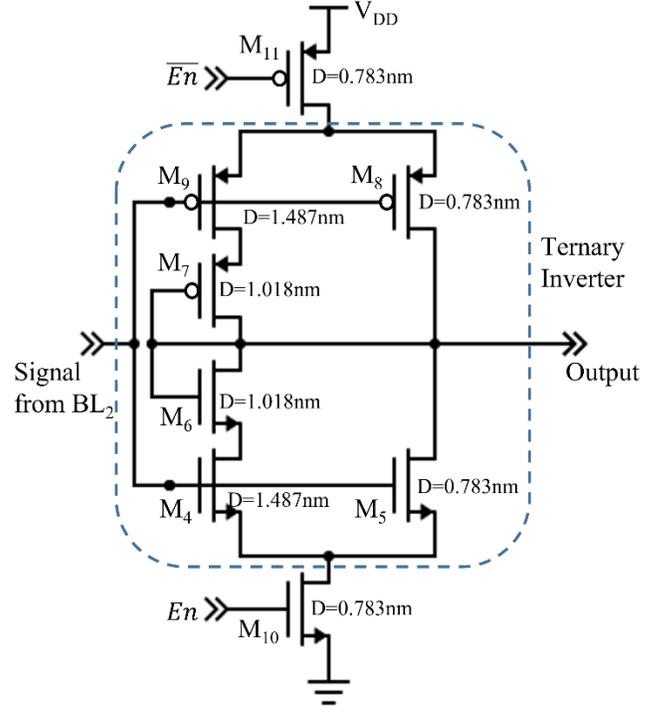

Fig. 9: Proposed Sense Circuitry

Table III: Truth Table for a Ternary Inverter

| Input | Output |
|---|---|
| 0 | 2 |
| 1 | 1 |
| 2 | 0 |

Two enable signals ($En$ and $\overline{En}$) are used for cutting the whole sense structure from power grid to save power. When $En = 1$, the circuit gets connected to the $V_{DD}$ and ground and works properly. During which, the read operation is sensed as follows:

- Case I (Read "0"):
  At this stage, the output at BL$_2$ is $V_{DD}$ (Fig. 5). As this signal is fed into the inverter of the sense circuit, the output obtained at the final output node is logic "0", according to the operating property of ternary inverter. No delay is present at this stage for reading "0".

- Case II (Read "1"):
  During this stage, the output at $BL_2$ shows a slow transition from $V_{DD}$ to $1/2V_{DD}$. When this input is passed to the sense circuit, after reaching a certain threshold voltage, this transition is picked up by the inverter and a voltage level equal to $1/2V_{DD}$ is obtained at the output. There is a certain delay to this phase which prohibits the sense circuit to consider any white noise or a sudden fluctuation to be a deliberate signal.

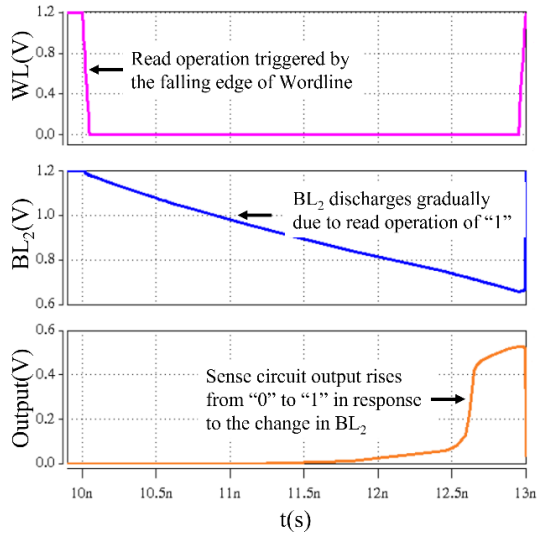

Fig. 10: Reading "1" from Sense Circuitry

- Case III (Read "2"):
  During reading "2" by the DRAM, the signal at the $BL_2$ shows a fast transition from $V_{DD}$ to ground. This sudden fall is sensed by the inverter of the sense circuit very swiftly and this low voltage level is converted to a high voltage level according to the property of ternary inverter.

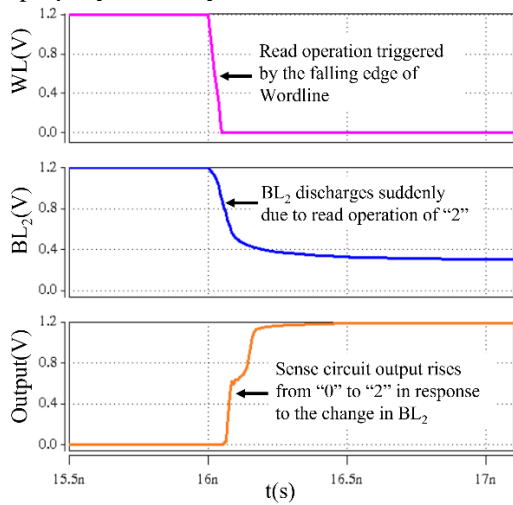

Fig. 11: Reading "2" from Sense Circuitry

The simulation results from the sense circuitry are given in Table IV. All the simulations are done in the presence of the system variables mentioned in Table II.

Table IV: Simulation results for the ternary 3T DRAM along with sense circuitry

| Transistor count for sense circuitry | Without Enable signal | 6 |
|---|---|---|
| | With Enable signal | 8 |
| Read sensing time for "0" | | 0ns |
| Read sensing time for "1" | | 2.0121ns |
| Read sensing time for "2" | | 0.083ns |
| Total Current from sense circuit | | 70.213nA |
| Total Power | | 84.2556nW |

## VI. CONCLUSION

In this paper, we have proposed a novel design for 3T ternary DRAM using DRAM. To the best of our knowledge, the proposed design is unique and has never been used. Here we have modified the binary DRAM to a ternary DRAM without the increase of any transistor. Besides that, we were able to decrease the number of read bit-line from two to one. We have also showed that the effect of other process variations is very less in the design, i.e. the design is very stable. Along with the proposed ternary DRAM, we have designed a sense circuitry containing enable signal which can be turned on and off completely from the main circuit according to need maintaining higher power efficiency.

REFERENCES

[1] Zahoor, Furqan, Tun Zainal Azni Zulkifli, Farooq Ahmad Khanday, and Sohiful Anuar Zainol Murad. "Carbon Nanotube and Resistive Random Access Memory Based Unbalanced Ternary Logic Gates and Basic Arithmetic Circuits." IEEE Access (2020).
[2] S. G. Hamedani and M. H. Moaiyeri, ``Impacts of process and temperature variations on the crosstalk effects in sub-10 nm multilayer graphene nanoribbon interconnects,'' *IEEE Trans. Device Mater. Rel.*, vol. 19, no. 4, pp. 630_641, Dec. 2019.
[3] Smith, Kenneth C. "The prospects for multi-valued logic: A technology and applications view." IEEE Transactions on Computers 9 (1981): 619-634.
[4] Donald E Knuth. The art of computer programming, vol 1: Fundamental. Algorithms. Reading, MA: Addison-Wesley, 1968.
[5] Anthony J McAuley and Paul Francis. Fast routing table lookup using cams. In IEEE INFOCOM'93 The Conference on
Computer Communications, Proceedings, pages 1382–1391. IEEE, 1993.
[6] Sergio R Ramirez-Chavez. Encoding don't cares in static and dynamic content-addressable memories. IEEE Transactions on Circuits and Systems II: Analog and Digital Signal Processing, 39(8):575–578, 1992
[7] Vincent Gaudet. A survey and tutorial on contemporary aspects of multiple-valued logic and its application to
microelectronic circuits. IEEE Journal on Emerging and Selected Topics in Circuits and Systems, 6(1):5–12, 2016.
[8] Sandhie ZT, Ahmed FU, Chowdhury MH. Design of ternary master-slave D-flip flop using MOS-GNRFET. In2020 IEEE 63rd International Midwest Symposium on Circuits and Systems (MWSCAS) 2020 Aug 9 (pp. 554-557). IEEE.
[9] Jaber RA, Kassem A, El-Hajj AM, El-Nimri LA, Haidar AM. High-

performance and energy-efficient CNFET-based designs for ternary logic circuits. IEEE Access. 2019 Jul 11;7:93871-86.

[10] Z. T. Sandhie, F. Uddin Ahmed and M. Chowdhury, "GNRFET based Ternary Logic – Prospects and Potential Implementation," 2020 IEEE 11th Latin American Symposium on Circuits & Systems (LASCAS), 2020, pp. 1-4, doi: 10.1109/LASCAS45839.2020.9069028.

[11] Z. T. Sandhie, F. U. Ahmed and M. H. Chowdhury, "Design of Ternary Logic and Arithmetic Circuits Using GNRFET," in IEEE Open Journal of Nanotechnology, vol. 1, pp. 77-87, 2020.

[12] M. U. Mohammed, R. Vijjapuram and M. H. Chowdhury, "Novel CNTFET and Memristor based Unbalanced Ternary Logic Gate," 2018 IEEE 61st International Midwest Symposium on Circuits and Systems (MWSCAS), 2018, pp. 1106-1109.

[13] Chen, Changxin, and Yafei Zhang. Nanowelded carbon nanotubes: From field-effect transistors to solar microcells. Springer Science & Business Media, 2009.

[14] Chang-Jian, Shiang-Kuo, Jeng-Rong Ho, and J-W. John Cheng. "Characterization of developing source/drain current of carbon nanotube field-effect transistors with n-doping by polyethylene imine." *Microelectronic Engineering* 87, no. 10 (2010): 1973-1977.

[15] C. Kocabas, S. Hur, A. Gaur, M. Meitl, M. Shim, and J. Rogers, "Guided growth of large-scale, horizontally aligned arrays of single-walled carbon nanotubes and their use in thin-film transistors," *Small*, vol. 1, pp. 1110–1116, 2001.

[16] C. Wang, K. Ryu, A. Badmaev, N. Patil, A. Lin, S. Mitra, H.-S. P. Wong, and C. Zhou, "Device study, chemical doping, and logic circuits based on transferred aligned single-walled carbon nanotubes," *Appl. Phys. Lett.*, vol. 93, no. 3, pp. 033101-1–033101-3, 2008.

[17] P. Zarkesh-Ha, A. Arabi, and M. Shahi, "Stochastic analysis and design guidelines for CNFETs in gigascale integrated systems," *IEEE Trans. Electron Devices*, vol. 58, no. 2, pp. 530–539, Feb. 2011.

[18] Deng J, Wong HS. A circuit-compatible SPICE model for enhancement mode carbon nanotube field effect transistors. In 2006 International Conference on Simulation of Semiconductor Processes and Devices 2006 Sep 6 (pp. 166-169). IEEE

[19] Carbon Nanotube Field Effect Transistor [Online]. Available: https://en.wikipedia.org/wiki/carbon_nanotube_fieldeffect_ transistor Accessed: March 23, 2020, 2018. URL https://en.wikipedia.org/wiki/Carbon_nanotube_field-effect_ transistor.

[20] Gyungseon Seol, Youngki Yoon, James K Fodor, Jing Guo, Akira Matsudaira, Diego Kienle, Gengchiau Liang, Gerhard Klimeck, Mark Lundstrom, Ahmed Ibrahim Saeed (2019), "CNTbands," https://nanohub.org/resources/cntbands-ext. (DOI: 10.21981/QT2F-0B32).

[21] B. Q. Wei, R. Vajtai, and P. M. Ajayan, 'Reliability and Current Carrying Capability of Carbon Nanotubes' Appl. Phys. Lett., vol. 79, pp. 1172-1174, 2001.

[22] Zarhoun R, Moaiyeri MH, Farahani SS, Navi K. An Efficient 5‐Input Exclusive‐OR Circuit Based on Carbon Nanotube FETs. ETRI Journal. 2014 Feb;36(1):89-98.

[23] Sandhie, Z.T., Patel, J.A., Ahmed, F.U. and Chowdhury, M.H., 2021. Investigation of Multiple-valued Logic Technologies for Beyond-binary Era. ACM Computing Surveys (CSUR), 54(1), pp.1-30.

[24] Lin, Sheng, Yong-Bin Kim, and Fabrizio Lombardi. "CNTFET-based design of ternary logic gates and arithmetic circuits." IEEE transactions on nanotechnology 10, no. 2 (2011): 217-225.

[25] Sandhie ZT, Patel JA, Ahmed FU, Chowdhury MH. Investigation of Multiple-valued Logic Technologies for Beyond-binary Era. ACM Computing Surveys (CSUR). 2021 Jan 21;54(1):1-30.